\documentclass{ifacconf}
\usepackage{amsmath}
\usepackage{amssymb}
\usepackage{relsize}
\usepackage{color}
\usepackage{mySymbols}
\usepackage{graphicx}
\usepackage[numbers, sort&compress, square]{natbib}
\usepackage{resizegather}
\newcommand*{\fiteq}[1]{\scalebox{3.4in}{1}}%
\newcommand{\tr}{\text{tr}}
\newcommand{\gap}{3.5pt}
\def\EXX{\mathbb{E}[x(k)x^T(k)]}
\begin{document}
\setlength{\abovedisplayskip}{3.2pt}
\setlength{\belowdisplayskip}{3.4pt}
\begin{frontmatter}
\title{On Simultaneous Long-Short Stock Trading Controllers with Cross-Coupling}
\author[madison]{Atul Deshpande} 
\author[madison]{John A. Gubner} 
\author[barmish]{B. Ross Barmish}

\address[madison]{Department of Electrical and Computer Engineering, \\University of Wisconsin, Madison, WI 53706.\\
email: atul.deshpande@wisc.edu; john.gubner@wisc.edu}%
\address[barmish]{Department of Electrical and Computer Engineering, \\
Boston University, Boston, MA 02215. \\email: bob.barmish@gmail.com}

\begin{abstract}
The \emph{Simultaneous Long-Short} (SLS) controller for trading a single stock is known to guarantee positive expected value of the resulting gain-loss function with respect to a large class of stock price dynamics. In the literature, this is known as the \emph{Robust Positive Expectation} (RPE) property.
An obvious way to extend this theory to the trading of two stocks is to trade each one of them using its own independent SLS controller.
Motivated by the fact that such a scheme does not exploit any correlation between the two stocks, we study the case when the relative sign between the drifts of the two stocks is known. 
The main contributions of this paper are three-fold: 
First, we put forward a novel architecture in which we cross-couple two SLS controllers for the two-stock case.
Second, we derive a closed-form expression for the expected value of the gain-loss function. 
Third, we use this closed-form expression to prove that the RPE property is guaranteed with respect to a large class of stock-price dynamics. 
When more information over and above the relative sign is assumed, additional benefits of the new architecture are seen. For example, when bounds or precise values for the means and covariances of the stock returns are included in the model, numerical simulations suggest that our new controller can achieve lower trading risk than a pair of decoupled SLS controllers for the same level of expected~trading~gain.
\end{abstract}
\begin{keyword}
Finance, Robustness, Stochastic Control, Uncertain Dynamic Systems.
\end{keyword}
\end{frontmatter}

\section{Introduction}
\label{sec:intro}
The starting point for this paper is the fact that the so-called {\it Simultaneous Long-Short\/} (SLS) stock trading controller, see \C{Barmish2011a,Barmish2016,Barmish2011,Baumann2017a,Malekpour2016,Deshpande2018}, guarantees satisfaction of the {\it Robust Positive Expectation\/} (RPE) property, which means the following: The trading gain-loss function is guaranteed to have positive expected value for a broad class of stock-price processes.
In this paper, we go beyond the single-stock results cited above and pursue this theme in a more general two-stock trading scenario. To this end, we introduce {\it Cross-Coupled\/} SLS controllers to exploit a ``correlation'' between the stocks. For this new controller, we prove an RPE theorem which holds when the sign of the cross-coupling coefficient is chosen appropriately.
When additional information about the price processes is assumed, our numerical examples suggest that cross-coupled SLS controllers can achieve lower trading risk than a pair of decoupled SLS controllers for the same target value of the expected trading gain.\\[\gap]
For an SLS controller trading a single stock, the key idea in existing literature involves generating the investment level in the stock from calculations based on hypothetically holding both long and short positions at the same time. This is accomplished using two complementary controllers.
The associated RPE results, obtained using linear feedback, differ from earlier work which is based on sample paths such as \cite{Dokuchaev2002a,Dokuchaev2002,Dokuchaev2004}, and model-specific trading strategies such as \C{markowitz1952,zhang2001stock,song2013optimal,deshpande2016general}.
In addition to the work in \C{Barmish2011a,Barmish2016,Barmish2011,Baumann2017a,Malekpour2016,Deshpande2018}, other contributions to the SLS theory involve robustness results with respect to stock prices having time-varying drift and volatility \C{Primbs2017}, prices generated from Merton's diffusion model \C{Baumann2017}, generalization to the case of PI controllers \C{Malekpour2013}, and discrete-time systems with delays \C{Malekpour2016}. 
More recently, in \C{Obrien2018}, the authors generalize the RPE Theorem to the case of an SLS controller which can have different parameters for the long and short sides of the trade and suggest procedures for controller parameter selection to minimize trading risk based on historical data.
In \C{Maroni2019}, the authors start with a problem formulation which treats prices as if they are disturbances, as in \C{Barmish2008}, and obtain the SLS controller parameters as the solution of an appropriately constructed \m{\mathcal{H}_{\infty}} optimization~problem.
 \\[\gap]
{A noticeable attribute of the SLS literature is that it addresses single-stock trading scenarios. 
For the multi-stock case, the obvious approach for using existing results would be to independently trade each stock using its own SLS controller, without exploiting any information about their price correlation.
Influenced by these considerations,  the innovation in \C{Deshpande2018} is to trade one stock with a long-only linear feedback and the other with a short-only linear feedback instead of using two separate SLS controllers. 
Additionally, a strong assumption on the price relationships between the two stocks is made.}
In contrast to the aforementioned, in this paper we impose a much weaker assumption. 
Specifically, we assume that only the relative sign between the underlying drifts of the two stock prices is known. 
We then put forward a new architecture that cross-couples two SLS controllers to take advantage of this relative-sign~information.\\[\gap]
In comparison to existing SLS literature, our new control architecture includes an extra degree of freedom: a cross-coupling feedback parameter \m{\coupling}, which forces interactions between the two SLS controllers.
In addition to the introduction of this novel trading architecture, the main theoretical contributions in this paper are results related to the expected value of the trading gain-loss function.
First, we provide a closed-form expression for the expected value of the trading gain-loss function.
Subsequently, we prove that for a range of \m{\coupling}, satisfaction of the RPE property is guaranteed with respect to a large class of stochastic processes for the stock prices. We also establish a recursive formula to calculate the variance of the gain-loss~function.\\[\gap]
Under strengthened assumptions that additional information about the stock prices is known over and above the relative sign between the mean returns, 
we demonstrate via a numerical simulation example that there can be performance benefits due to the use of  cross-coupling.
In our example, using an assumed price model with known means and variances of the stock-price returns, an optimized cross-coupled architecture achieves lower risk than two similarly optimized independent SLS controllers. 
The methodology used in the numerical example can be easily adapted to evaluate performance benefits when the price model assumed is not precise. 
For example, when the drifts and volatilities of the price processes are characterized with bounds instead of precise values, minimax optimization of the controller design is still possible.\\[\gap]
Additionally, in our numerical example, we consider the account leverage resulting from the use of the cross-coupled controller. 
For the specific case of Geometric Brownian Motion prices, along most sample paths, trading with the optimal cross-coupled controller results in lower account leverage than that obtained with optimal independent SLS controllers. For scenarios with a limit on the trading account leverage, we see that a ``saturated'' implementation of our cross-coupled controller still results in trading gains with a positive sample mean.

\section{Two-Stock Trading Scenario}
\label{setup}
{In this section, we describe our two-stock trading setup, including the assumptions which are in force.
\subsubsection*{Stock Price Dynamics}
We consider two stocks with stochastically varying prices \m{S_1(k)} and \m{S_2(k)} having associated returns 
$$
\rho_1(k)\doteq \frac{S_1(k+1) - S_1(k)}{S_1(k)};\;\;\rho_2(k)\doteq \frac{S_2(k+1) - S_2(k)}{S_2(k)}
$$
at stages \m{k=0,1,\dots, N-1}, with the assumption that
the return vectors~$[\rho_1(k)\;\;\rho_2(k)]^T$ are independent and identically distributed. The mean values of the returns~$
\mu_1\doteq \mathbb{E}[\rho_1(k)]$ and $\mu_2\doteq\mathbb{E}[\rho_2(k)]
$ 
are unknown to the trader. Only the relative sign between the two means, namely
\m{\text{sign}(\mu_1\mu_2)},
is assumed to be known.
\if 0
, where \m{\text{sign}(\cdot)} is defined as
$$
\text{sign}(x) = 
\begin{cases}
 &1,\;\;\;\;\;  x>0;\\
 &0,\;\;\;\;\; x=0;\\
 -\hspace{-.3cm}&1,\;\;\;\;\; x<0.
\end{cases}
$$
\fi
For instance, if the two stocks are from the same sector, it is often the case that they tend to move in the same direction; i.e., \m{{\rm sign}(\mu_1\mu_2)=1}, over the medium to long term; e.g., see \C{King1966}.
Similarly, when assets in a portfolio are negatively correlated; e.g., see \C{Irwin87}, we assume \m{{\rm sign}(\mu_1\mu_2)=-1}.
\subsubsection*{Idealized Market Assumptions}
In the theory to follow, consistent with existing SLS literature, an idealized market is assumed.
That is, transaction costs such as brokerage, commissions, taxes, or fees levied by the stock exchange, are not incurred.
In addition, we assume perfect liquidity so that there is no gap between the bid and ask prices, and the trader can buy or sell any number of shares, including fractions, at the market price.
These assumptions are similar to those made in the context of ``frictionless markets'' in finance literature; e.g., see \cite{merton1992continuous}.
\subsubsection*{Leverage and Interest} 
In practice, the broker usually imposes limits on the trading account leverage. 
For our theoretical analysis, however, we assume that leverage limits are not in play.
That is, the trader has sufficient account resources to hold any desired position in the stocks. 
In Section \R{sec:simulations}, when we provide a numerical example, we study the practical implications of a leverage constraint and suggest further research on this issue in Section \R{conclusion}.
We also assume that the margin interest and the risk-free rates of return are zero; we defer consideration of nonzero rates to~future~research.
%Despite the above assumptions, note that we study the practical implications of controller design choice on the trading account leverage in Section \R{sec:simulations}.
}

%\section{Cross-Coupled LS Controller}
\section{Two Independent SLS Controllers}\label{sec:2sls_setup}
{
To provide context for the analysis and main results to follow, we first elaborate on the obvious way that existing  single-stock SLS theory might be applied to the two-stock case. 
As discussed in Section 1, one can simply design two decoupled SLS controllers: one for the first stock and another for the second. 
Proceeding in this manner, the net investment levels \m{I_1(k)} and\m{I_2(k)} in the stocks at stage $k$ are obtained as sums
$$
I_1(k) = I_{_{1L}}(k)+I_{_{1S}}(k);\;\;\;I_2(k) = I_{_{2L}}(k)+I_{_{2S}}(k),
$$
where \m{I_{_{iL}}(k)} and \m{I_{_{iS}}(k)} for \m{i=1,2,} are the nominally long and short positions in the $i$-th stock, each obtained using a linear feedback controller.
That is, with initial investment levels \m{I_{01}>0} and \m{I_{02}>0} and feedback parameters \m{K_1>0} and \m{K_2>0}, the long and short investment~functions are given respectively~as
\begin{align*}
I_{iL}(k) = I_{0i} + K_i g_{_{iL}}(k);\;\;\;
I_{iS}(k) = - I_{0i} - K_i g_{_{iS}}(k),
\end{align*}
for \m{i=1,2}, where the cumulative gain-loss functions resulting from individual long and short positions in each stock are obtained using the gain-loss update equations
\begin{align*}
g_{_{iL}}(k+1) &= g_{_{iL}}(k) + I_{_{iL}}(k)\rho_i(k);\\
g_{_{iS}}(k+1) &= g_{_{iS}}(k) + I_{_{iS}}(k)\rho_i(k)
\end{align*}
for \m{i=1,2},
with \m{g_{_{1L}}(0)=g_{_{1S}}(0)=g_{_{2L}}(0)=g_{_{2S}}(0)=0}.
In the sequel, we refer to the above as the 2-SLS controller.\\[\gap]
Now, the overall trading gain-loss function \m{g(k)} for this setup is given~by
$$
g(k) \doteq g_{_{1L}}(k) + g_{_{1S}}(k) + g_{_{2L}}(k) +g_{_{2S}}(k).
$$
Applying existing results as in \C{Deshpande2018} to each of the two stocks individually, we arrive at
\if 0
Since these linear feedbacks constitute two independent single-stock SLS controllers that individually have the RPE property (see appendix of \C{Deshpande2018}), this property is extended to the two-stock setup by default. In fact, it is straightforward to obtain the closed-form expression for the resulting expected gain-loss function as
\fi
$$ 
\Exp{g(N)} = \sum_{i=1}^{2}{\frac{I_{0i}}{K_i}\left[(1+K_i\mu_i)^N+(1-K_i\mu_i)^N-2\right]},
$$
which is positive for \m{N>1}, and \m{\mu_1}, \m{\mu_2} not both~zero.}
\section{Cross-Coupling and State Equations}\label{sec:ccsls_setup}
In this section, we describe the main technical novelty of this paper: a new architecture for trading which involves cross-coupling between two single-stock SLS controllers. 
This is achieved by augmenting each of the four linear investment functions of Section \R{sec:2sls_setup} with a coupling term having feedback gain \m{-1 < \coupling < 1} to obtain
\begin{align*}
I_{iL}(k) &= I_{0i} + K_i g_{_{iL}}(k) - \coupling K_j g_{_{jS}}(k);\\
I_{iS}(k) &= - I_{0i} - K_i g_{_{iS}}(k) + \coupling K_j g_{_{jL}}(k).
\end{align*}
for \m{i,j\in\{1,2\}};\m{i\neq j}.
We refer to this as the Cross-Coupled SLS (CC-SLS) controller.
When $\coupling=0$, we recover the decoupled 2-SLS~controller and the corresponding results stated in~Section \R{sec:2sls_setup}.\\[\gap]
To provide some insight into the operation of the CC-SLS controller, we consider the case when the drifts \m{\mu_1} and \m{\mu_2} as well as \m{\coupling} are positive. For this case, we expect \m{g_{_{1 L}}(k)} and \m{g_{_{2 L}}(k)} to both be positive, with \m{g_{_{1 S}}(k)} and \m{g_{_{2 S}}(k)} both negative. 
%which would be ``expected'' since both drifts are positive, 
This leads to \m{I_{1 L}(k)} and \m{I_{2 L}(k)} being greater than would be the case if \m{\coupling=0}; i.e., greater than the long-investment levels in the 2-SLS counterpart. 
Another case which can be similarly analyzed is encountered when the first stock follows a sample path with a positive trend, but the second stock does not, so that \m{g_{2 S}(k)} is positive; resulting in a smaller \m{I_{1 L}(k)} than would be the case without coupling. More generally, the CC-SLS controller invests more aggressively or less aggressively than its 2-SLS counterpart, depending on the extent to which the stock behaviors are consistent with their drifts.
\subsubsection*{Gain-Loss Update Equations}
Substituting the formulae for the relevant CC-SLS investment functions into each of the four gain-loss update equations in Section \R{sec:2sls_setup}, we obtain the closed-loop~equations
\begin{gather*}
{g_{_{{iL}}}(k+1)} = (1+K_i\rho_i(k)){g_{_{iL}}(k)} -\coupling K_j\rho_i(k) {g_{_{jS}}(k)}+ I_{_{0i}}\rho_i(k);\\[\gap]
{g_{_{iS}}(k+1)} = (1-K_i\rho_i(k)){g_{_{iS}}(k)} +\coupling K_j\rho_i(k) {g_{_{jL}}(k)}- I_{_{0i}}\rho_i(k)
\end{gather*}
for \m{i,j\in\{1,2\}};\m{i\neq j}, with overall trading gain-loss function 
$
g(k)= g_{_{1L}}(k)+g_{_{1S}}(k)+g_{_{2L}}(k)+g_{_{2S}}(k).
$
\subsubsection*{State-Space Representation}
We work with a state-space representation of the CC-SLS controller to derive our main results on robust positivity of \m{\Exp{g(N)}}.
%Reordering and rewriting the gain-loss update equations in vector state-space form, with
With~state
$$
{x}(k) \doteq  \begin{bmatrix}
 {g_{_{1L}}(k)}&
 {g_{_{2S}}(k)}&
 {g_{_{1S}}(k)}&
 {g_{_{2L}}(k)}
 \end{bmatrix}^T
$$
and \m{c \doteq [1\;\;1\;\;1\;\;1]^T}, the output of interest is
$$
g(k) = c^Tx(k).
$$
Then, the gain-loss update equations become
\begin{align*}
{x}(k+1) &= A(k){x}(k)+{b}(k){u}(k);\\[\gap]
{y}(k) &\doteq {c}^T{x}(k)
\end{align*}
where
\begin{gather*}
A(k) = \begin{bmatrix}
1+K_1\rho_1(k) & -\coupling K_2 \rho_1(k) & 0 & 0\\
\coupling K_1\rho_2(k) & 1-K_2\rho_2(k) &0&0\\
0&0&1-K_1\rho_1(k) & \coupling K_2 \rho_1(k) &\\
0&0&-\coupling K_1\rho_2(k) & 1 + K_2 \rho_2(k)\\
\end{bmatrix},\\[\gap]
{b}(k) = \begin{bmatrix}
I_{_{01}}\rho_1(k)&
-I_{_{02}}\rho_2(k)&
-I_{_{01}}\rho_1(k)&
I_{_{02}}\rho_2(k)
\end{bmatrix}^T,
\end{gather*}
and constant input \m{{u}(k)\equiv 1.}
Since \m{x(0)=\mathbf{0}}, the standard solution for \m{{g(N)}} is
\begin{align*}
{g(N)}={y}(N)= {c}^T\sum_{k=0}^{N-1}\Phi(N,k+1) {b}(k) {u}(k),
\end{align*}
where the state transition matrix $\Phi(k,k_0)$ from stage \m{k_0} to \m{k\geq k_0} is given~by
$$
\Phi(k,k_0) \doteq 
\begin{cases}
                      A(k-1)\cdots A(k_0+1)\cdot A(k_0) &\text{ for } k>k_0;\\
                      I_{4\times 4} &\text{ for } k=k_0.
                     \end{cases}
$$
{
Taking expectations, we obtain
\begin{align*}
\Exp{g(N)}={c}^T\sum_{k=0}^{N-1} \barA^{N-1-k} {\barb},
\end{align*}
where
\begin{gather*}
 \barA \doteq \Exp{A(k)} = \begin{bmatrix}
1+K_1\mu_1 & -\coupling K_2 \mu_1 & 0 & 0\\
\coupling K_1\mu_2 & 1-K_2\mu_2 &0&0\\
0&0&1-K_1\mu_1 & \coupling K_2 \mu_1 &\\
0&0&-\coupling K_1\mu_2 & 1 + K_2 \mu_2\\
\end{bmatrix}
\end{gather*}
and 
$$
\barb \doteq \Exp{b(k)} = \begin{bmatrix}
I_{_{01}}\mu_1 &
-I_{_{02}}\mu_2 &
-I_{_{01}}\mu_1&
I_{_{02}}\mu_2
\end{bmatrix}^T.\\[-10pt]
$$
}\label{sec:statespace}
\section{Main Results}\label{sec:rpe}
In this section, we provide two theorems related to the expected value \m{\Exp{g(N)}} of the overall gain-loss function at stage \m{N}. 
The first theorem gives us the formula for \m{\Exp{g(N)}}. Following this, the second theorem gives us conditions under which \m{\Exp{g(N)}>0}. 
For simplicity of the proofs, we assume that both \m{\mu_1} and \m{\mu_2} are nonzero.  However, by separately considering the case when one of these two drifts vanish, it is easy to see that \m{\Exp{g(N)}>0} in this situation, except for the break-even case when both \m{\mu_1} and \m{\mu_2} are zero.
%Note that in Section \R{sec:ccsls_setup}, we already addressed the degenerate cases, when either \m{\mu_1}, \m{\mu_2} or both vanish, and the special case of 2-SLS obtained when \m{\coupling=0}.
\\[\gap]
To obtain a formula for \m{\Exp{g(N)}}, we use the following notation. For \m{K_1>0}, \m{K_2>0}, and \m{0<|\coupling|<1}, we~define
\begin{align*}
 \theta&\doteq\sqrt{(K_1\mu_1+K_2\mu_2)^2-4\coupling^2K_1K_2\mu_1\mu_2},\\
 \alpha_1\ &\doteq (\theta-K_1\mu_1+K_2\mu_2)/2,\\
\alpha_2\ &\doteq (\theta+K_1\mu_1-K_2\mu_2)/2,\\
\beta_1 &\doteq(K_1\mu_1 + K_2\mu_2 + \theta),\\
 \beta_2 &\doteq (K_1\mu_1 + K_2\mu_2 - \theta),
\end{align*}
and the function
$$
\phi_{_N}(x) \doteq  (1 + x)^N+ (1 - x)^N-2,
$$
which is positive for all \m{x\neq 0} and \m{N>1}.
\subsubsection*{Expected Value Theorem}
\emph{Suppose two stocks with stochastically varying prices \m{S_1(k)} and \m{S_2(k)} with mean returns \m{\mu_1} and \m{\mu_2} are traded using the Cross-Coupled SLS controller with \m{K_1>0}, \m{K_2>0}, and coupling coefficient satisfying \m{0<|\coupling|<1}. Then, for \m{\mu_1} and \m{\mu_2} nonzero, the expected value of the gain-loss function is given~by}
\begin{gather*}
\Exp{g(N)} = \frac{1}{2\theta}\Bigg[
\frac{2\coupling \mu_1\mu_2  (I_{_{01}}K_1+ I_{_{02}}K_2)+I_{_{02}}\mu_2\beta_1 + I_{_{01}}\mu_1\beta_2}{\alpha_1} \cdot{\phi_{_N}(\alpha_1)}\\
\qquad + \frac{2\coupling \mu_1\mu_2  (I_{_{01}}K_1+ I_{_{02}}K_2)+I_{_{02}}\mu_2\beta_2 + I_{_{01}}\mu_1\beta_1
}{\alpha_2}\cdot{\phi_{_N}(\alpha_2)}
\Bigg].
\end{gather*}
%\subsubsection*{Derivation of the Formula} 
%Here we derive the formula for \m{\Exp{g(N)}} described above.
\subsubsection*{Proof}
For \m{\mu_1,\mu_2\neq 0} and coupling coefficient \m{0<|\coupling|<1}, we diagonalize the block-diagonal matrix \m{\barA} defined in Section \R{sec:ccsls_setup} to obtain
$$
\barA = P\Lambda P^{-1},
$$
where
$$
P\doteq\begin{bmatrix}
\frac{\beta_2}{2\coupling K_1\mu_2} & \frac{\beta_1}{2\coupling K_1\mu_2} & 0 & 0\\
1 & 1  & 0 & 0\\
0 & 0 & \frac{\beta_1}{2\coupling K_1\mu_2} & \frac{\beta_2}{2\coupling K_1\mu_2}\\
0& 0& 1 & 1
\end{bmatrix}
$$
is composed of the eigenvectors of \m{\barA} as its columns,
and
$$
\Lambda\doteq\diag\left(
1 - \alpha_1, \;\; 1 + \alpha_2, \;\; 1 - \alpha_2, \;\; 1 + \alpha_1\right)
$$
is the diagonal matrix with the corresponding eigenvalues of \m{\barA}.
Note that the standing assumptions assure that \m{\alpha_1} and \m{\alpha_2} are nonzero.\\[\gap]
Rewriting the expression for \m{\Exp{g(N)}} in terms of \m{P} and \m{\Lambda}, we obtain
\begin{align*}
\Exp{g(N)}&= c^T\sum_{k=0}^{N-1} \barA^{N-1-k} \barb\\
&=c^T\sum_{i=0}^{N-1}\left( P\Lambda^{N-1-i} P^{-1}\right)\barb\\
%&= c^T P\left(\sum_{i=0}^{N-1} \Lambda^{N-1-i}\right) P^{-1}\barb\\
&=c^T P\Lambda_{_S} P^{-1}\barb,
\end{align*}
where the diagonal matrix \m{\Lambda_{_S}}
%is the eigenvalue matrix of \m{\displaystyle\sum_{k=0}^{N-1} \barA^{N-1-k}} and 
is given by
\if 0
\begin{gather*}
 \Lambda_{_S}=\begin{bmatrix}
\frac{(1 - \alpha_1)^N-1}{-\alpha_1} & 0 & 0 & 0\\
0 & \frac{(1 + \alpha_2)^N-1}{\alpha_2} & 0 & 0\\
0 & 0 & \frac{(1 - \alpha_2)^N-1}{-\alpha_2} & 0\\
0& 0& 0 & \frac{(1 + \alpha_1)^N-1}{\alpha_1}
\end{bmatrix}.
\end{gather*}
\fi
\begin{gather*}
 \Lambda_{_S}\doteq\diag\left(
\frac{(1 - \alpha_1)^N-1}{-\alpha_1}, \;\; \frac{(1 + \alpha_2)^N-1}{\alpha_2}, \;\; \frac{(1 - \alpha_2)^N-1}{-\alpha_2}, \;\; \frac{(1 + \alpha_1)^N-1}{\alpha_1}
\right).
\end{gather*}
Since \m{\Exp{g(N)}} is a scalar, we write
$$
\Exp{g(N)} = c^T P\Lambda_{_S} P^{-1}\barb = \tr\left( c^T P\Lambda_{_S} P^{-1}\barb \right),
$$
where \m{\tr(\cdot)} is the trace operator, and its cyclic property~\C{strang09} gives~us
\begin{align*}
 \Exp{g(N)} &= \tr(c^T P\Lambda_{_S} P^{-1}\barb) \\
 &=\tr(P\Lambda_{_S} P^{-1}\barb c^T) \\
 &= \tr(\Lambda_{_S} P^{-1}\barb c^TP) \\
 &= \sum_{i}(\Lambda_{_S})_{ii} (P^{-1}\barb c^TP)_{ii},
\end{align*}
where \m{(\Lambda_{_S})_{ii}} and \m{(P^{-1}\barb c^TP)_{ii}} denote the diagonal entries of the respective matrices. For \m{\Lambda_{_S}} as given above, we need only find the values of \m{(P^{-1}\barb c^TP)_{ii}},
%the last step resulting from \m{\Lambda_S} being a diagonal matrix.
\if 0
Finally, square matrices $X$ and $Y$, we know that
$$
\tr(XY^T) = \sum_{i,j} X_{ij} Y_{ij}.
$$ 
Since \m{\Lambda_{_S}} is a diagonal matrix,  to compute this inner product
\fi
%Thus, we only need the diagonal elements of \m{P^{-1}\barb c^TP}, which we obtain upon simplification~as 
which we collect in the vector
\begin{align*}
 D &\doteq \begin{bmatrix}(P^{-1} \barb c^T P)_{11}\\ 
 (P^{-1} \barb c^T P)_{22}\\
 (P^{-1} \barb c^T P)_{33}\\
 (P^{-1} \barb c^T P)_{44}
 \end{bmatrix}.
\end{align*}
Further simplification using \m{\beta_1\beta_2 = 4\coupling^2 K_1 K_2\mu_1\mu_2} yields
$$
%\text{diag}(P^{-1} \barb c^T P)
D = \frac{1}{2\theta} \begin{bmatrix}
-I_{_{02}}\mu_2(\beta_1 + 2\coupling K_2\mu_1) - I_{_{01}}\mu_1(\beta_2 + 2\coupling K_1\mu_2)\\
I_{_{02}}\mu_2(\beta_2 + 2\coupling K_2\mu_1) + I_{_{01}}\mu_1(\beta_1 + 2\coupling K_1\mu_2)\\
-I_{_{02}}\mu_2(\beta_2 + 2\coupling K_2\mu_1) - I_{_{01}}\mu_1(\beta_1 + 2\coupling K_1\mu_2)\\
I_{_{02}}\mu_2(\beta_1 + 2\coupling K_2\mu_1) + I_{_{01}}\mu_1(\beta_2 + 2\coupling K_1\mu_2)
\end{bmatrix}.
$$
Then the summation above for \m{\Exp{g(N)}} simplifies to the claimed closed-form expression. $\qed$
\subsubsection*{Remarks}
We observe that the expected value of the gain-loss function is of the form 
$$
\Exp{g(N)} = 
C_1\cdot\phi_{_N}(\alpha_1)
 + C_2\cdot\phi_{_N}(\alpha_2),
$$
where \m{C_1} and \m{C_2} are independent of \m{N}.
This is similar in form to the result given for two independent SLS controllers in Section \R{sec:2sls_setup}, that is,
$$
\Exp{g(N)}\Bigl|_{\mbox{\footnotesize{2-SLS}}} = \frac{I_{_{01}}}{K_1} \phi_N(K_1\mu_1) + \frac{I_{_{02}}}{K_2} \phi_N(K_2\mu_2).
$$
\if 0
Furthermore, observing that for small \m{x}, 
$
\phi(x)\approx N(N-1)x^2,
$
and using the formulas
\begin{align*}
 \alpha_1\beta_1+\alpha_2\beta_2 
&= 2K_2\mu_2\theta;\\[\gap]
 \alpha_1\beta_2+\alpha_2\beta_1 &= 2K_1\mu_1\theta,
 \end{align*}
 it follows that for suitably small \m{K_1|\mu_1|}, \m{K_2|\mu_2|}, the CC-SLS closed-form expression reduces~to
 \begin{align*}
\Exp{g(N)} 
&\approx
{N(N-1)}\Bigl[
 I_{_{01}}K_1\mu_1^2 + I_{_{02}}K_2\mu_2^2\Bigl]\\ 
&\qquad+ 
{N(N-1)}\coupling\mu_1\mu_2(I_{_{01}}K_1+ I_{_{02}}K_2).
\end{align*}
The first term above is an approximation of the expected gain-loss function of the 2-SLS controller, which is positive for \m{N>1}.
Furthermore, when we choose \m{\coupling} so that ${\rm sign}(\coupling)= {\rm sign}(\mu_1\mu_2),$
the second term is also positive.
Thus, if the 2-SLS controller with \m{\coupling=0} has the other parameters \m{I_{_{01}}, I_{_{02}}, K_1}, \m{K_2}, the same as those used for the CC-SLS controller, the result is an approximation of \m{\Exp{g(N)}} which is strictly smaller than that of the CC-SLS controller.
\fi 
In the theorem to follow, we use the closed-form expression for \m{\Exp{g(N)}} to prove that if~$
{\rm sign}(\coupling)= {\rm sign}(\mu_1\mu_2),$
the RPE property of the CC-SLS controller is guaranteed.
\if 0
\section{Gain-Loss Dynamics at High-Frequency}
\input{gainlossdynamics}
\fi
%\section{Main Result}\label{sec:rpe}
%\begin{samepage}
\subsubsection*{Robust Positive Expectation Theorem}
\emph{Suppose two stocks with stochastically varying prices \m{S_1(k)} and \m{S_2(k)} with nonzero mean returns \m{\mu_1} and \m{\mu_2} are traded using the Cross-Coupled SLS controller with coupling coefficient satisfying \m{0<|\coupling|<1} and \m{{\rm sign}(\coupling)={\rm sign}(\mu_1\mu_2)}.
Then, for any~${N>1}$, robust satisfaction of the condition
$$
\Exp{g(N)}>0
$$
is guaranteed.}
%\end{samepage}
\begin{table*}[!hbtp]
\centering
{
 \begin{tabular}{|c|c|c|c|c|c|}
 \hline
Scenario  
 & Bounds on $\theta$ & ${\alpha_1}$ & ${\alpha_2}$ & ${\beta_1}$ & ${\beta_2}$\\
 \hline
 ${\mu_1>0;\mu_2>0}$ & { ${|K_1\mu_1 -K_2\mu_2|<\theta< K_1\mu_1+K_2\mu_2}$} & ${\alpha_1>0}$ & ${\alpha_2>0}$ & ${\beta_1>0}$ & ${\beta_2>0}$ \\
 \hline
 ${\mu_1<0;\mu_2<0}$ & {${|K_1\mu_1 -K_2\mu_2|<\theta< |K_1\mu_1+K_2\mu_2|}$} & ${\alpha_1>0}$ & ${\alpha_2>0}$ & ${\beta_1<0}$ & ${\beta_2<0}$ \\
 \hline
 ${\mu_1>0;\mu_2<0}$ & {${|K_1\mu_1 +K_2\mu_2|<\theta< K_1\mu_1-K_2\mu_2}$} & ${\alpha_1<0}$ & ${\alpha_2>0}$ & ${\beta_1>0}$ & ${\beta_2<0}$ \\
 \hline
 ${\mu_1<0;\mu_2>0}$ & {${|K_1\mu_1 +K_2\mu_2|<\theta< K_2\mu_2-K_1\mu_1}$} & ${\alpha_1>0}$ & ${\alpha_2<0}$ & ${\beta_1>0}$ & ${\beta_2<0}$ \\
 \hline
 \end{tabular}}
 \vspace{5pt}
 \caption{Satisfaction of the First Two Inequalities for All Combinations of Signs of \m{\mu_1} and \m{\mu_2}}
 \label{mutable}
\end{table*}

\if 0
\section{Proof of the Theorem}\label{sec:thmproof}
\fi
\subsubsection*{Proof}
{
\if 0
For fixed \m{I_{_{01}}, I_{_{02}}, K_1, K_2}, all positive, and \m{0<|\coupling|<1}, we prove that the expected value of the gain-loss function, \m{\Exp{g(N)}}, of the cross-coupled SLS architecture described in Section~\ref{sec:ccsls_setup} is positive for an arbitrary pair \m{\mu_1\neq 0} and \m{\mu_2\neq 0}
such that
$$
\text{sign}(\mu_1\mu_2) = \text{sign}(\coupling).
$$
%\FloatBarrier
\fi
Beginning with the formula obtained for \m{\Exp{g(N)}} and rearranging, it suffices to show that the following three inequalities~hold:
\begin{align*}
\frac{\phi_{_N}(\alpha_1)\cdot (I_{_{02}}\mu_2\beta_1 + I_{_{01}}\mu_1\beta_2)}{\alpha_1}&>0;\\
 \frac{\phi_{_N}(\alpha_2)\cdot (I_{_{02}}\mu_2\beta_2 + I_{_{01}}\mu_1\beta_1
)}{\alpha_2}&>0;\\
2\coupling \mu_1\mu_2  (I_{_{01}}K_1+ I_{_{02}}K_2) \Bigg(\frac{\phi_{_N}(\alpha_1)}{\alpha_1}+
\frac{\phi_{_N}(\alpha_2)}{\alpha_2}\Bigg)&\geq 0.
\end{align*}
For arbitrary admissible pair \m{\mu_1\neq 0} and \m{\mu_2\neq 0}, we verify the satisfaction of the first two inequalities above for the cases enumerated in Table \R{mutable}. In each row of the table, we consider a possible combination of the signs of \m{\mu_1} and \m{\mu_2}. Each combination determines the range of possible values \m{\theta} can take, which in turn dictates the signs of \m{\alpha_1, \alpha_2,\beta_1} and \m{\beta_2}. Using these in conjunction with the positivity of \m{\phi_N(\alpha_i)} for \m{N>1} establishes the first two inequalities.\\[\gap]
To prove the third inequality holds, given that 
$$
\text{sign}(\mu_1\mu_2)=\text{sign}(\coupling),
$$ 
we readily see that
$2\coupling \mu_1\mu_2 (I_{_{01}}K_1 + I_{_{02}}K_2)>0.$
Thus, to complete the proof, it suffices to show that
$$
\frac{\phi_{_N}(\alpha_1)}{\alpha_1}+
\frac{\phi_{_N}(\alpha_2)}{\alpha_2}\geq 0.
$$
To this end, we consider the following two cases:\\[\gap]
{\it Case 1:} If \m{\text{sign}(\mu_1)=\text{sign}(\mu_2)}, we see from Table \R{mutable} that \m{\alpha_1>0} and \m{\alpha_2>0}. Combined with the fact that the function \m{\phi_N(x)>0} for all \m{x\neq 0}, it follows that \m{\phi_N(\alpha_1)>0} and \m{\phi_N(\alpha_2)>0}. Hence,
$$
\frac{\phi_{_N}(\alpha_1)}{\alpha_1}+
\frac{\phi_{_N}(\alpha_2)}{\alpha_2}>0.\\
$$
{\it Case 2:} If \m{\text{sign}(\mu_1)=-\text{sign}(\mu_2)}, we see from Table \R{mutable} that \m{\text{sign}(\alpha_1)=-\text{sign}(\alpha_2)}. Without loss of generality, assuming \m{\mu_1>0>\mu_2}, we obtain \m{\alpha_2>0>\alpha_1} with \m{|\alpha_2|\geq|\alpha_1|} and we use this condition to arrive at
$$
\frac{\phi_{_N}(\alpha_1)}{\alpha_1}+\frac{\phi_{_N}(\alpha_2)}{\alpha_2}
=\frac{|\alpha_1|\phi_{_N}(\alpha_2)-|\alpha_2|\phi_{_N}(\alpha_1)}{|\alpha_1\alpha_2|}.
$$
Since \m{N>1}, it is easily shown that 
$$
|\alpha_1|\phi_{_N}(\alpha_2)>|\alpha_2|\phi_{_N}(\alpha_1).
$$
\if 0
From the binomial theorem for \m{N>1}, we readily obtain
$$
\phi_N(x) = 2\sum_{k = 1}^{\lfloor N/2 \rfloor} {N \choose 2k} x^{2k}.
$$
Substituting this above, we can rewrite 
$$
\frac{\phi_{_N}(\alpha_1)}{\alpha_1}+\frac{\phi_{_N}(\alpha_2)}{\alpha_2}=2\sum_{k = 1}^{\lfloor N/2 \rfloor} \frac{{N \choose 2k} (|\alpha_1|\alpha_2^{2k}-|\alpha_2|\alpha_1^{2k})}{|\alpha_1\alpha_2|}\geq 0.
$$
\fi
%proving that 
%$$
%\frac{\phi_{_N}(\alpha_1)}{\alpha_1}+
%\frac{\phi_{_N}(\alpha_2)}{\alpha_2}\geq 0.
%$$
%The proof for \m{\mu_2>0>\mu_1} is obtained in a similar manner.
This completes the proof of the theorem.
%The two scenarios show that Condition \R{cond1} is satisfied for the arbitrary chosen pair \m{\mu_1,\mu_2}.
%The satisfaction of Conditions (\ref{cond1}), (\ref{cond2}) and (\ref{cond3}) proves part~{\bf (b)} of the theorem. 
\m{\qed}}

\section{Variance of the Gain-Loss Function}\label{sec:variance}
{
Assuming that the mean vector
$$
\mu\doteq [\mu_1\;\; \mu_2]^T\doteq \Exp{\rho_1(k)\;\; \rho_2(k)}^T,
$$
and covariance matrix
$$
{\sigma} \doteq \begin{bmatrix}
 \sigma_1^2 & \sigma_{12}\\
 \sigma_{12}&\sigma_2^2\\
\end{bmatrix}\doteq {\rm cov}([\rho_1(k)\;\; \rho_2(k)]^T)
$$
of the returns of the two stocks are known,
we now derive a recursion to calculate \m{\var(g(N))}.
Recalling the state-update~equation 
from Section \R{sec:statespace}, we first rewrite the state matrix as
$$
A(k) = A_0\rho_0(k) + A_1\rho_1(k) + A_2\rho_2(k)
$$
where \m{\rho_0(k)\equiv 1},  $A_0$ is the identity matrix \m{I}, 
\begin{gather*}
A_1 \doteq \begin{bmatrix}
K_1 & -\coupling K_2 & 0 & 0\\
 0 & 0 &0&0\\
0&0&-K_1 & \coupling K_2 &\\
0&0& 0 & 0\\
\end{bmatrix},\;\;A_2 \doteq \begin{bmatrix}
0 & 0 & 0 & 0\\
\coupling K_1 & -K_2 &0&0\\
0&0&0 & 0 &\\
0&0&-\coupling K_1 & K_2\\
\end{bmatrix},
\end{gather*}
and we rewrite \m{b(k)} as
$$
b(k) = b_0\rho_0(k) +b_1\rho_1(k) +b_2\rho_2(k),
$$
where 
\begin{gather*}
b_0 \doteq [0\;\;0\;\;0\;\;0]^T;
b_1 \doteq [I_{_{01}}\;\;0\;\;-I_{_{01}}\;\;0]^T;
b_2 \doteq [0\;\;-I_{_{02}}\;\;0\;\;I_{_{02}}]^T.
\end{gather*}
With this notation, the state update equation~becomes
\begin{gather*}
 x(k+1) = (A_0\rho_0(k)+A_1\rho_1(k)+A_2\rho_2(k))x(k)\\ 
 \qquad\qquad\qquad+ (b_0\rho_0(k)+b_1\rho_1(k)+b_2\rho_2(k)),
\end{gather*}
with initial value \m{x(0) = {0}}. It follows that
$$
\mathbb{E}[x(k+1)] = \barA \mathbb{E}[x(k)] + \barb.
$$
To calculate the variance of the gain-loss function, we first 
define \m{\rho(k) \doteq [\rho_0(k)\;\; \rho_1(k)\;\; \rho_2(k)]^T}, and subsequently
\begin{align*}
 R(\mu,{\sigma}) &\doteq \Exp{\rho(k)\rho(k)^T}\\
&= \begin{bmatrix}
1 & \mu_1 & \mu_2 \\
 \mu_1 & \sigma_1^2+\mu_1^2 &\sigma_{12} + \mu_1\mu_2 \\
\mu_2 & \sigma_{12} + \mu_1\mu_2 & \sigma_2^2+\mu_2^2\\
\end{bmatrix}.
\end{align*}
We now obtain the recursion
\begin{gather*}
\mathbb{E}[x(k+1)x^T(k+1)] = \mathbb{E}[(A(k)x(k) + b(k))(A(k)x(k) + b(k))^T],
\end{gather*}
Substituting the various quantities into the recursion above for \m{\EXX} and denoting the ($i,j$)-th element of the matrix \m{R(\mu,{\sigma})} as \m{R_{i,j}(\mu,{\sigma})}, it follows that
\begin{gather*}
 \mathbb{E}[x(k+1)x^T(k+1)] = \sum_{i=0}^2\sum_{j=0}^2 R_{i+1,j+1}(\mu,{\sigma})\Bigl[A_i\EXX A_j^T\\ 
 \qquad\qquad\qquad+ A_i\mathbb{E}[x(k)] b_j^T + b_i\mathbb{E}[x^T(k)] A_j^T + b_ib_j^T\Bigl].
\end{gather*}
Starting with initial value 
${\Exp{x(0)x^T(0)}=0}$, we use the above recursion for \m{0\leq k \leq N-1} to obtain \m{\mathbb{E}[x(N)x^T(N)]}.
Recalling that the gain-loss function ${g(N)=c^Tx(N)}$ for \m{c = [1\;\;1\;\;1\;\;1]^T}, 
we now calculate
\begin{align*}
\Exp{g(N)} = c^T \Exp{x(N)};\;\;
\Exp{g^2(N)} = c^T \Exp{x(N)x(N)^T} c,
\end{align*}
and subsequently,
$$
\var(g(N)) = \Exp{g^2(N)} - \mathbb{E}^2[g(N)].
$$
From this, we calculate the standard deviation of \m{g(N)}, which is used in the numerical example in Section \R{sec:simulations}.}
\if 0
\subsubsection*{Remark}
Although the terms introduced above involving \m{b_0} are zero, this enables a more compact representation of the final result for the variance above.
\fi

\section{Risk Mitigation via Cross-Coupling}\label{sec:risk}
{
To augment the analysis of the CC-SLS controller in Section \R{sec:ccsls_setup}, we 
now analyze a trading scenario where a cross-coupling can result in lower trading risk when compared to two independent SLS controllers.
This analysis is performed under the strengthened assumption that the mean returns and covariances of the two stock prices are known.
%given a target gain, we can come up with a CC-SLS controller design that achieves lower trading risk than two independent SLS controllers.
%we study the effect of cross-coupling on the mean and variance of the gain-loss function, with the standard deviation of the gain-loss function representing the trading risk, 
Then in the spirit of modern portfolio theory, for the classical case when the controller parameters \m{I_{_{01}}, I_{_{02}}, K_1, K_2} and \m{\coupling} are optimized with respect to these assumed price models,
we compare the mean and standard deviation of \m{g(N)} obtained by the CC-SLS controller against the ones obtained by the 2-SLS controller.
%That is, we consider the case when more information about the stochastic prices is assumed, and 
%we optimize the controller parameters \m{I_{_{01}}, I_{_{02}}, K_1, K_2} and \m{\coupling} to achieve an acceptable balance between the .
%With this consideration in mind, in Section \R{sec:simulations},  
We demonstrate how, for a given target return, the CC-SLS controller architecture can lead to lower trading risk than that of the 2-SLS controller. Namely, the standard deviation of \m{g(N)} resulting from the use of the CC-SLS is lower than that obtained using~2-SLS.\\[\gap]
Associated with the two controllers described in Sections~\ref{sec:2sls_setup} and \ref{sec:ccsls_setup}, whenever convenient,
we emphasize the dependence of various quantities on the controller parameter~vector
$$
d\doteq {(I_{_{01}},I_{_{02}},K_1,K_2,\coupling)}
$$ 
by including it as an argument in mathematical functions of interest; e.g., we write \m{g(d,k)} instead~of \m{g(k)} for the gain-loss function at stage \m{k}. 
Thus, given initial trading account value \m{V_0}, the account value \m{V(k)} at stage \m{k} is given~by
$$
V(d,k) = V_0 + g(d,k)
$$
and depends on \m{d}.
Without loss of generality, we assume \m{V_0=1} so that the cumulative return
$$
 \frac{V(d,N)-V_0}{V_0}
$$
is equal to \m{g(d,N)}, and the risk-return pair~is
$$
\bigl(\std(g(d,N)),{\Exp{g(d,N)}}\bigr).\\[-13pt]
$$}
\section{Numerical Example}\label{sec:simulations}
%\subsubsection*{CC-SLS and 2-SLS Controller Parameters}
To illustrate the ideas in Section \R{sec:risk}, we work with an assumed stochastic model of the stock-price processes with independent returns having respective mean values~${\mu_1=0.023374}$ and ${\mu_2=0.031014},
$ 
and known variances~$
{\sigma_1^2=8.3333\times 10^{-3}}$ and ${\sigma_2^2=16.333\times 10^{-3}}.
$
%For this model, we calculate the expected value and variance of the gain-loss function for 5000 CC-SLS controller designs picked uniformly at random from the range
Then, for both the CC-SLS and the 2-SLS controllers, we consider 5000 candidate parameter vectors~$d$ selected by using the uniform distribution to generate~points
\begin{align*}
 I_{_{01}},I_{_{02}}\in(0, 3];\;\;
 K_1,K_2\in(0,3];\;\;
 \coupling \in [0,0.99],
\end{align*}
noting that \m{\coupling=1} is inadmissible in the RPE Theorem.
For each vector \m{d} selected above for the CC-SLS controller, we force \m{\coupling=0} to get a corresponding 2-SLS parameter vector.
Then for each \m{d}, and \m{N=30}, we calculate the risk-return pair~$\left({\std(g(d,N))}, \Exp{g(d,N)}\right)$
for each of the two controllers. In Figure \R{var}, for the CC-SLS controller, each such pair is denoted with a blue dot and for the 2-SLS controller, a green dot is~used.
\\[\gap]
\begin{figure}[h]
\centering
\includegraphics[trim = 0cm 0cm 0cm 0cm,clip, width=3.4in]{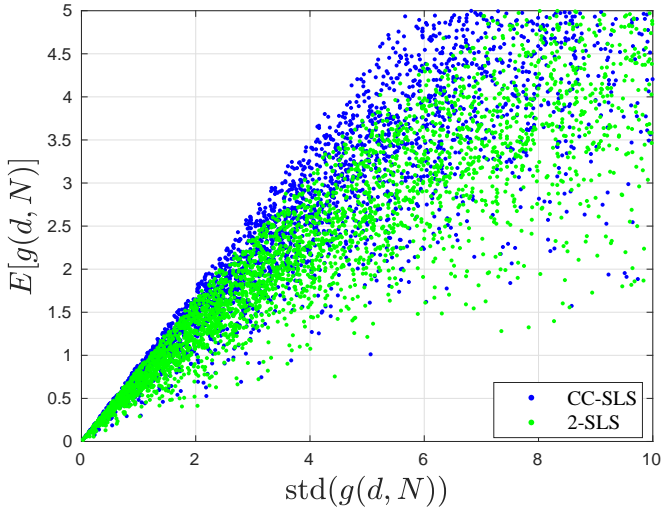} 
\caption{Expected Value vs. Standard Deviation of \m{g(d,N)}}
\label{var}
\end{figure}
For each of the two architectures, for a given target return \m{G}, we seek a parameter vector \m{d} that~solves
\begin{gather*}
\min_{d}\std(g(d,N)) \text{ subject to } \Exp{g(d,N)}\geq G. 
\end{gather*}
In terms of Figure \R{var}, this means that the CC-SLS optimum parameter vector corresponds to the leftmost blue dot along the line with
$
\Exp{g(d,N)}\approx G.
$
Similarly for 2-SLS, the optimum parameter vector corresponds to the leftmost green dot along the same line.
From the figure, we see that for target return \m{G}, the optimal CC-SLS controller appears to guarantee a lower level of the risk \m{\std(g(d,N))} versus that obtained for the optimal 2-SLS controller.\\[\gap]
To make the above more concrete, for \m{G=2}, using the search sets for the controller parameters above, %using the respective leftmost dots as the initial values of the controller designs, we 
%solve the design optimization problem for each controller in MATLAB. 
%We find that 
an optimal 2-SLS controller parameter vector is found to~be
$$
d^*_{2sls} =  (3, 3, 0.713, 0.381)
$$
and results in 
$$
\left(\std(g(d^*_{2sls},N)), \Exp{g(d^*_{2sls},N)}\right) \approx (2.399,2.00).
$$
{Similarly, an optimal CC-SLS parameter vector is found to be }
$$
d^*_{ccsls} = (3, 2.58, 0.339, 0.234,0.990),
$$
and results in
$$
\left(\std(g(d^*_{ccsls},N)), \Exp{g(d^*_{ccsls},N)}\right) \approx    (2.225,2.00).
$$
Consistent with the discussion above, this CC-SLS controller leads to lower risk than its 2-SLS counterpart.
{
\subsubsection*{Account Leverage Considerations}
As stated in Section \R{setup}, the broker typically imposes limits on the account leverage to ensure that investment levels are commensurate with the account value. 
Thus, to supplement the foregoing risk-return analysis, we study the leverage used by the two optimal controllers above.
Given that one or both stocks may be sold short with the corresponding net investments \m{I_i(k)<0}, consistent with practice, we work with leverage ratio
%For one million price sample paths, with \m{I_i(k)} denoting the amount at risk in the \m{i}-th stock and the account leverage at stage \m{k} 
$$
L(k) \doteq \frac{|I_{_1}(k)|+|I_{_2}(k)|}{V(k)},
$$
and study
$
L_{\max} \doteq \displaystyle\max_{0\leq k\leq N-1} L(k)
$
using one million sample paths. 
For simulating these paths, however, we need to know the joint distribution of the returns, not just their means and covariances. To this end, we assume that the prices are obtained from two independent Geometric Brownian Motion models which are consistent with the \m{(\mu_i,\sigma_i^2)} used for the two controller optimization tasks above. Since the total returns \m{S_i(k+1)/S_i(k)} are log-normally distributed, we generate these prices using the update equations %Then, working again with \m{N=30}, calculating
%, we obtain price sample paths using the update~equations
\begin{align*}
 S_1(k+1) &= S_1(k)\cdot \exp\left(0.019142+0.08903 w_1(k)\right);\\
 S_2(k+1) &= S_2(k)\cdot \exp\left(0.022918+0.12349 w_2(k)\right),
\end{align*}
where \m{w_1(k)\sim\mathcal{N}(0,1)} and \m{w_2(k)\sim\mathcal{N}(0,1)}. Without loss of generality, we assume that the initial prices in the above update equations are  \m{S_1(0)=S_2(0)=1}.
Subsequently, for \m{N=30}, we estimate that for the CC-SLS controller,~$
L_{\max}\leq 3.44
$ 
about 95\% of the time,
and that for the 2-SLS controller, using that same 95\% figure of merit, we estimate 
$
{L}_{\max}\leq 6.94.
$ 
Furthermore, out of the one million sample paths, the optimal CC-SLS controller results in a bankruptcy, characterized by account value \m{V(k)\leq 0}, in only $715$ sample paths as compared to $15,006$ bankruptcies for the 2-SLS controller. %Bankruptcy in this case is characterized by the account value going to zero or in the negative, and 
For such sample paths, we record the maximum leverage to be \m{L_{\max}=\infty.}
To summarize, the optimal CC-SLS controller not only leads to lower risk than its optimal 2-SLS controller, but it also results in a much lower account leverage almost all the time and leads to a lower probability of account bankruptcy.}
\subsubsection*{Saturated 2-SLS and CC-SLS Controllers}
Noting that the account leverage for both controllers can far exceed the limits imposed by stock brokers, to illustrate how one might conform with common practice, we revisit our simulation with the added constraint \m{L(k)\leq 2}, and ``saturate'' the investments of the 2-SLS and the CC-SLS controllers whenever this inequality is violated. More precisely, we take
$$
I_{_{iL}}(k) = \begin{cases}
I_{_{iL}}(k) &\text{ when } L(k)\leq 2;\\
I_{_{iL}}(k)\cdot 2/L(k) &\text{ otherwise},
\end{cases}
$$
and
$$
I_{_{iS}}(k) = \begin{cases}
I_{_{iS}}(k) &\text{ when } L(k)\leq 2;\\
I_{_{iS}}(k)\cdot 2 /L(k) &\text{ otherwise}.
\end{cases}
$$
Although such a scheme ensures that \m{L(k)\leq 2} for all \m{k}, the theoretical guarantee of robust positive expectation is no longer available. Nonetheless, from the one million sample paths of the GBM prices described above, we {\it statistically estimate} \m{\Exp{g(N)}\approx 1.86} and \m{{\std}({g}(N))\approx 2.15}, and encounter no bankruptcies using the saturated CC-SLS controller. In comparison, for the saturated 2-SLS controller, we estimate \m{\Exp{g(N)}\approx 1.74}, with \m{{\std}({g}(N))=2.27} and face account bankruptcy in 12 sample paths.
Thus, for this more practical scenario, the saturated CC-SLS and 2-SLS controllers yield positive average trading gains while conforming to the leverage constraints imposed on them.
\section{Conclusion}\label{conclusion}
In this paper, we introduced the notion of cross-coupled SLS controllers for trading two stocks.
%We present a novel cross-coupled SLS (CC-SLS) controller setup for trading two stocks when the relative sign between their underlying price drifts is known, incorporating this information into the controller design via the coupling coefficient.
We derived a closed-form expression for the expected trading gain-loss function resulting from this new architecture and, based on this formula, our new Robust Positive Expectation Theorem provides conditions under which positivity of the expected gain-loss function is guaranteed.
We also provided simulations which suggest, under strengthened hypotheses, that our new CC-SLS architecture enables controller designs that achieve lower trading risk than two independent SLS controllers for the same target expected gain.
{Finally, in our numerical simulations, we found that the cross-coupled SLS controller results in lower trading account leverage than two decoupled SLS controllers. Similarly, we showed that using ``saturated'' implementations of both controller designs which guarantee compliance with a leverage limit imposed by the broker, the mean-variance performance of the cross-coupled SLS controller is better than that obtained with two decoupled SLS controllers.}\\[\gap]
Three directions for future research immediately present themselves: 
The first involves the introduction of cross-coupling into different variants of the SLS controller found in the literature; e.g., in \C{Malekpour2016}, an SLS controller with delays is considered. A second direction involves extending the theory presented in this paper to trading scenarios involving more than two stocks.
Finally, the third possible direction for future research is motivated by the fact that the practitioner invariably faces leverage restrictions. Hence, it would be of interest to see if the CC-SLS controller still leads to a guarantee of robust positive expectation when a saturation scheme, along the lines studied in the previous section, is used. Results given in \C{Barmish2016} for a standalone SLS controller provide motivation for further~work.\\[-20pt]

% Generated by IEEEtran.bst, version: 1.14 (2015/08/26)
% Generated by IEEEtran.bst, version: 1.14 (2015/08/26)
% Generated by IEEEtran.bst, version: 1.14 (2015/08/26)
\bibliographystyle{plainnat}
\bibliography{ccsls}
\if 0

\fi
\end{document}